\documentclass[%
 aip,
 amsmath,amssymb,
 longbibliography,
 reprint,%
]{revtex4-2}

\usepackage{graphicx}
\usepackage{dcolumn}
\usepackage{bm}

\usepackage[utf8]{inputenc}
\usepackage[T1]{fontenc}
\usepackage{mathptmx}
\usepackage{xcolor}
\usepackage{soul}
\usepackage{color}
\usepackage{multirow}\usepackage{xspace}
\usepackage{amsmath}
\usepackage{amsfonts}
\usepackage{amssymb}
\usepackage{amsthm,bm}
\usepackage{commath}
\usepackage{setspace}

\usepackage{titlesec}

\titleformat{\paragraph}[runin]
  {\normalfont\normalsize\bfseries\itshape}{(\theparagraph)}{1em}{}
\titlespacing*{\paragraph}
  {0pt}{*}{0.5em}

\titleformat{\subparagraph}[runin]
  {\normalfont\normalsize\itshape}{(\thesubparagraph)}{1em}{}
\titlespacing*{\subparagraph}
  {0pt}{*}{0.5em}

\renewcommand\theparagraph{\roman{paragraph}}
\renewcommand\thesubparagraph{\alph{subparagraph}}

\newcommand{\bk}{{\bf k}}

\newcommand{\be}{\begin{equation}}
\newcommand{\ee}{\end{equation}}
\newcommand{\bea}{\begin{eqnarray}}
\newcommand{\eea}{\end{eqnarray}}
\newcommand{\beqa}{\begin{eqnarray*}}
\newcommand{\eeqa}{\end{eqnarray*}}

\newcommand{\ba}{\begin{array}{c}}
\newcommand{\baa}{\begin{array}{cc}}
\newcommand{\baaa}{\begin{array}{ccc}}
\newcommand{\baaaa}{\begin{array}{cccc}}
\newcommand{\ea}{\end{array}}

\newcommand{\bma}{\left[\begin{array}{c}}
\newcommand{\bmaa}{\left[\begin{array}{cc}}
\newcommand{\bmaaa}{\left[\begin{array}{ccc}}
\newcommand{\bmaaaa}{\left[\begin{array}{cccc}}
\newcommand{\ema}{\end{array}\right]}

\begin{document}

\preprint{AIP/123-QED}

\title{Tuning the Electronic and Optical Properties of Impurity-Engineered Two-Dimensional Graphullerene Half-Semiconductors}

\author{M. A. Khan}
\affiliation{NanoScience Technology Center, University of Central Florida, Orlando, FL 32826, USA.}
\author{$^{,2}$ Madeeha Atif}
\affiliation{Department of Physics, Federal Urdu University of Arts, Sciences and Technology, Islamabad, Pakistan.}
\author{Michael N. Leuenberger}
\affiliation{NanoScience Technology Center, Department of Physics, College of Optics and Photonics, University of Central Florida, Orlando, Fl 32826, USA.}



\begin{abstract}
A novel material consisting of a monolayer of C$_{60}$ buckyballs with hexagonal symmetry has recently been observed experimentally, named graphullerene. In this study, we present a comprehensive \textit{ab-initio} theoretical analysis of the electronic and optical properties of both pristine and impurity-engineered monolayer graphullerene using spin-dependent density functional theory (spin-DFT). Our findings reveal that graphullerene is a direct band gap semiconductor with a band gap of approximately 1.5 eV at the $\Gamma$ point, agreeing well with experimental data. Notably, we demonstrate that by adding impurities, in particular substitutional nitrogen, substitutional boron, or adsorbent hydrogen, to graphullerene results in the formation of spin-dependent deep donor and deep acceptor levels, thereby giving rise to a variety of half-semiconductors. All the impurities exhibit a magnetic moment of approximately $\mu_B$ per impurity. This impurity engineering enables the tuning of spin-polarized exciton properties in graphullerene, with spin-dependent band gap energies ranging from 0.43 eV ($\lambda \sim$ 2.9 $\mu$m) to 1.5 eV ($\lambda \sim$ 820 nm), covering the near-infrared (NIR) and short-wavelength infrared (SWIR) regimes. Our results suggest that both pristine and impurity-engineered graphullerene have significant potential for the development of carbon-based 2D semiconductor spintronic and opto-spintronic devices.
\end{abstract}


\maketitle
\begin{figure*}[hbt]
	\begin{center}
		\includegraphics[width=7in]{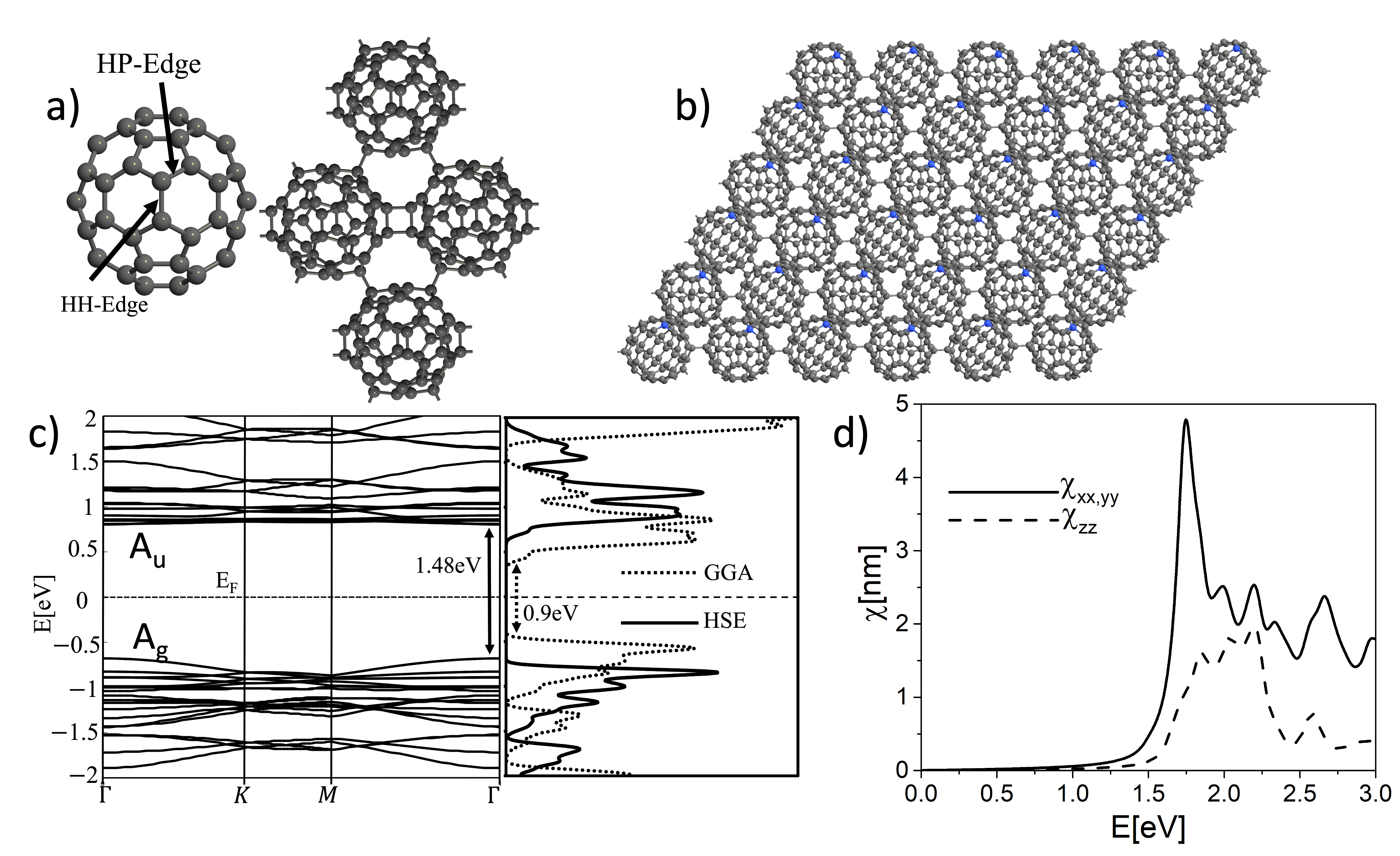}
	\end{center}
	\caption{a) Isolated C$_{60}$ molecule and graphullerene unit cell. b) Impurity-engineered graphullerene, dark grey (blue) color denotes carbon (impurity) atoms. c) Bandstructure and density of states of pristine graphullerene. The GGA calculation underestimates the band gap by a significant margin, with an error of 42$\%$ compared to the experimental value of 1.55eV,\cite{Hou2022} while the HSE functional gives a better value for the band gap with an error of 4.5$\%$. Therefore we use the HSE functional in our calculations. (d) Electric susceptibility of pristine graphullerene.}
	\label{structure_band_DOS}
\end{figure*} 

\section{Introduction}
Fullerenes are a class of molecules composed entirely of carbon atoms arranged in spherical or ellipsoidal cages. The most notable and extensively studied fullerene is C$_{60}$, also known as buckminsterfullerene, which resembles a soccer ball with carbon atoms at the vertices of a truncated icosahedron. Discovered in 1985, fullerenes have been the focus of extensive research due to their unique physical and chemical properties. They are highly stable and insoluble in most solvents, making them promising materials for various applications. In a groundbreaking experiment by Hou et al., a 2D crystal of C$_{60}$ molecules covalently bonded to each other was synthesized with quasi-hexagonal lattice symmetry using Mg helper atoms, denoted as qHPC$_{60}$.\cite{Hou2022} Utilizing the published qHPC$_{60}$ lattice structure, a density functional theory (DFT) study explored the electronic, optical, and mechanical properties of qHPC$_{60}$.\cite{Tromer2022} The band structure analysis revealed that the broken hexagonal symmetry necessitates a square lattice symmetry with high-symmetry points $\Gamma=(0,0)$, $X=(\frac{1}{2},0)$, $Y=(0,\frac{1}{2})$, and $U=(\frac{1}{2},\frac{1}{2})$. The unit cell comprises two C$_{60}$ molecules. Using the SIESTA code, a direct band gap of $E_g=0.9$ eV at the $\Gamma$-point was obtained. Due to the underestimation of the experimentally observed band gap of $E_{g,exp}=1.55$ eV,\cite{Hou2022} a scissor correction was applied to shift the absorption edge to 1.5-1.6 eV. The calculated optical response showed strong anisotropy between the $X$, $Y$, and $Z$ directions. Recent theoretical studies have demonstrated the thermal stability and fracture patterns of qHPC$_{60}$ through reactive molecular dynamics simulations,\cite{Ribeiro2022} as well as its stability and elasticity through electron localization function, crystal orbital Hamiltonian population, and density of states (DOS) extracted from DFT calculations.\cite{Shen2023} The anisotropic elastic and tensile properties of qHPC$_{60}$ have also been investigated.\cite{Zhao2023}

In a recent experiment, Meirzadeh et al. created a similar 2D material made of covalently bonded C$_{60}$ molecules.\cite{Meirzadeh2023} While they also used Mg helper atoms to synthesize this 2D material, they were able to remove the Mg atoms completely. The result is a 2D C$_{60}$ crystal with hexagonal symmetry, which they call graphullerene. Creating a 2D network of fullerenes involves the arrangement of C$_{60}$ molecules in a regular pattern, akin to how atoms are arranged in a crystal lattice. This can result in unique electronic, mechanical, and optical properties due to the couplings between the C$_{60}$ molecules and the specific arrangement of carbon atoms. 

Here we show that by adding substitutional nitrogen (N), substitutional boron (B), or adsorbent hydrogen (H) to graphullerene, it is possible to create a variety of half-semiconductors, which we classify by means of charge- and spin-dependent band gaps and spin-flip gaps. Our spin-dependent density functional theory (spin-DFT) calculations show that graphullerene is a direct band gap semiconductor with a spin-independent band gap of approximately 1.5 eV at the $\Gamma$ point, agreeing well with experimental data. In contrast to that, the impurity-engineered graphullerene half-semiconductors exhibit spin-polarized spin-dependent band gap energies ranging from 0.43 eV ($\lambda \sim$ 2.9 $\mu$m) to 1.5 eV ($\lambda \sim$ 820 nm), covering the near-infrared (NIR) and short-wavelength infrared (SWIR) regimes. Our results suggest that impurity-engineered graphullerene half-semiconductors could be useful for the development of spintronic and opto-spintronic devices based on deep impurity levels.

A C$_{60}$ molecule contains twelve pentagons and twenty hexagons, with each of the carbon atoms three-fold coordinated and three of the four valence electrons occupying two-center $\sigma$-like bonding orbitals separated from their unoccupied antibonding counterparts by an energy gap of more than 15 eV, starting 4 eV below the Fermi level. The large energy gap between the bonding and antibonding orbitals is characteristic of a highly stable molecule and contributes to the molecule's overall stability and resistance to reactivity, making C$_{60}$ a relatively inert molecule under ambient conditions. Many of the electronic and optical properties of the C$_{60}$ molecule can be explained using $\pi$-like orbitals ($p_z$), formed from 60 atomic orbitals (AO's) pointing in a radial direction. The eigenstates transform according to the irreducible representations (irreps) of the icosahedral symmetry of C$_{60}$. It is well known that the optical transition between the HOMO and LUMO states of the C$_{60}$ molecule is optically forbidden. In stark contrast to that, our DFT calculations reveal that for graphullerene the optical transition between the valence and conduction bands is allowed. Our results can have significant implications for understanding the electronic and optical properties of this novel material.

It has been shown that the optical band gap of C$_{60}$ can be tuned by impurity engineering, which involves introducing foreign atoms such as boron (B) and/or nitrogen (N) into the C$_{60}$ structure.\cite{TOPC60} Tuning the optical band gaps of materials is an active area of research with a wide range of promising applications such as in displays, nanoscale electronics, laser technology, optical filters, optical storage, scintillators, medical imaging detectors, optical sensors, optical switches, and solar cells. Here, we address the effects of impurity engineering on the electronic and optical properties of graphullerene by introducing N and B atoms as substituents. Besides the substitutional impurities N and B, we find that adsorbent H also creates localized states at low concentrations and impurity bands at high concentrations within the band gap of graphullerene. Thus, the band gap can be tuned in graphullerene by a suitable choice and concentration of impurities.

Interestingly, while pristine monolayer (ML) graphullerene is a regular semiconductor, spin-dependent density functional theory (spin-DFT) calculations reveal strongly spin-dependent band structures in all versions of impurity-engineered graphullerene. In other words, all versions of impurity-engineered ML graphullerene are half-semiconductors with spin-dependent band gaps. For a classification of half-metals and half-semiconductors, see for example Ref.~\onlinecite{Li2016}. Throughout this paper, we will study ML graphullerene. Therefore, we will omit the specification ML. We find that the band gap of one spin species can be two to three times larger than the band gap of the opposite spin species. This phenomenon provides the possibility to use non-polarized light to generate spin-polarized excitons by tuning the energy of the incident photons above the band gap of one spin species but below the band gap of the opposite spin species. We use group theory and representation theory to determine the selection rules for the optical transitions in pristine and impurity-engineered graphullerene. In the optical spectra, we clearly identify the large difference in optical band gaps between the two spin species.

\section{Electronic properties of pristine and impurity-engineered graphullerene}
\subsection{Pristine graphullerene}
\paragraph*{Structural properties:}
All numerical calculations are carried out by using DFT and with the use of hybrid GGA (HSE) as implemented in the Synopsis Atomistix Toolkit (ATK) 2022.\cite{QuantumATK} In a hybrid GGA calculation a fraction of Hartree-Fock exchange is mixed with the GGA exchange-correlation functional. This mixing parameter is typically denoted as "$\alpha$" and can vary between 0 and 1. When "$\alpha$" is 0, the functional reduces to standard GGA, and when "$\alpha$" is 1, it becomes pure Hartree-Fock. In our DFT calculations, we choose standard HSE06 hybrid functional parameters. The introduction of Hartree-Fock exchange in hybrid GGA calculations improves the description of certain properties, such as band gaps, which are often underestimated by standard GGA calculations. We consider a unit cell containing four C$_{60}$ molecules consisting of 240 carbon atoms. We add a vacuum separation of 25\AA\, between the graphullerene layers to suppress the coupling between adjacent graphullerene layers. The two pairs of C$_{60}$ molecules in the graphullerene unit cell have different orientations, i.e the C$_{60}$ molecules are rotated by 180$^\circ$ along the y-axis compared to neighboring C$_{60}$ molecules. The graphullerene structure is first geometrically optimized up to a force tolerance of 0.01 eV/\AA.   

Let us first describe the structural properties of graphullerene. Graphullerene is a molecular crystal in which the so-called intramolecular $\sigma$-bondings between the carbon atoms inside each C$_{60}$ molecule are stronger than the intermolecular $\pi$-bondings between few atoms of adjacent C$_{60}$ molecules.  
The lengths of the shared hexagon-hexagon (HH) (Fig: \ref{structure_band_DOS}(a)) and shared hexagon-pentagon (HP) (Fig: \ref{structure_band_DOS}(a)) edges within the C$_{60}$ molecules away from the intermolecular bondings are obtained as 1.400 \AA, and 1.446 \AA, respectively. These values are in good agreement with previously reported values for an isolated C$_{60}$ molecule.
\begin{figure*}[hbt]
	\begin{center}
		\includegraphics[width=7in]{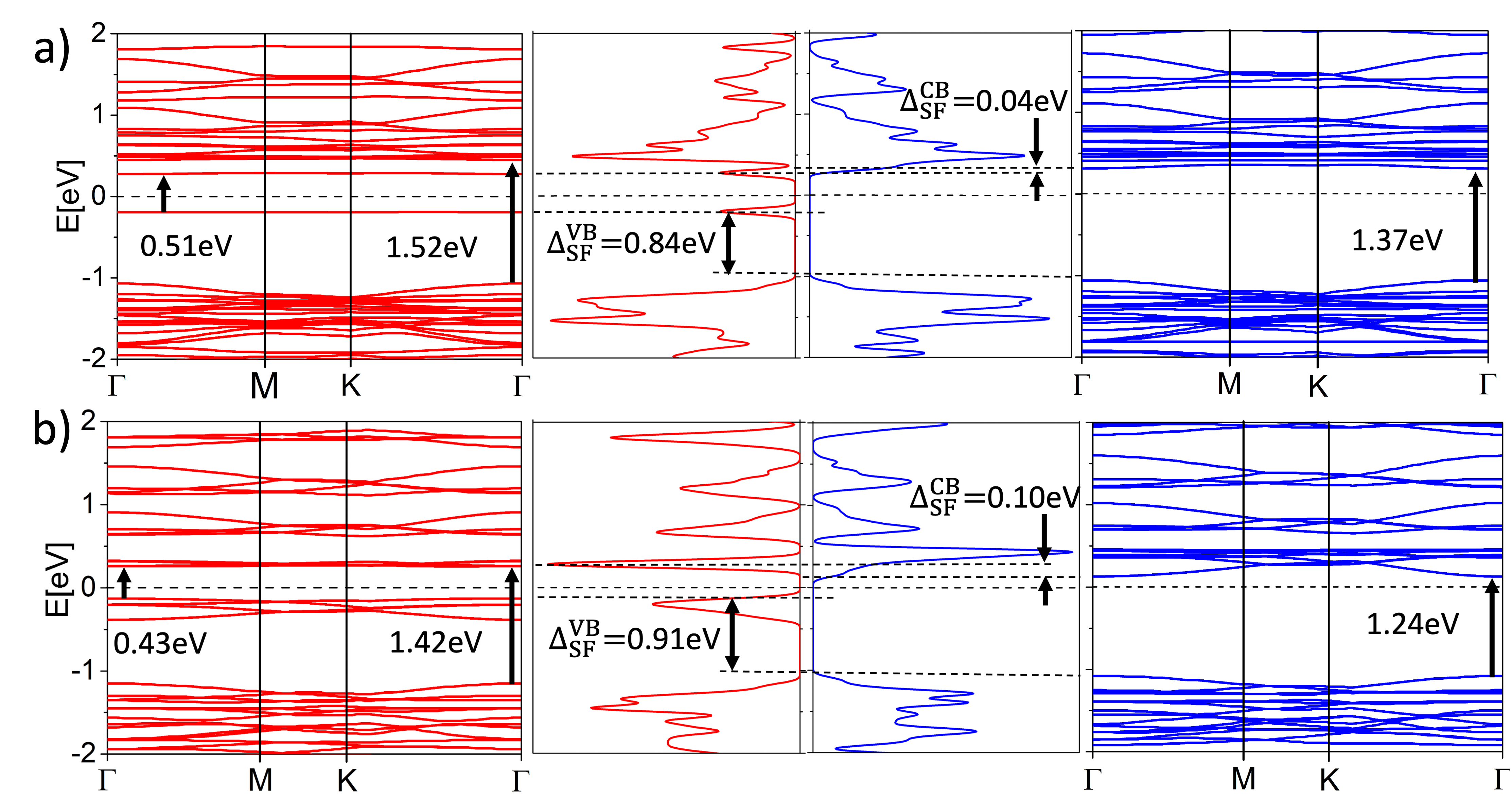}
	\end{center}
	\caption{a) The bandstructure and density of states for magnetic C$_{239}$N using spin-DFT. Red (blue) color is for spin up (down) electronic states. The charge, spin up, and spin down band gaps are $E_g=0.51$ eV, $E_{g\uparrow}=0.51$ eV, and $E_{g\downarrow}=1.37$ eV, respectively. The spin flip gaps in the VB and CB are $\Delta^{VB}_{SF}=0.84$ eV and $\Delta^{CB}_{SF}=0.04$ eV, respectively. The band gaps between the intrinsic electronic states for spin up and spin down are $E^{in}_{g\uparrow}=$ 1.52 eV and $E^{in}_{g\downarrow}=$ 1.37 eV, respectively. b) The bandstructure and density of states for magnetic C$_{236}$N$_4$ using spin-DFT. Red (blue) color is for spin up (down) electronic states. Red (blue) color is for spin up (down) electronic states. The charge, spin up, and spin down band gaps are $E_g=0.32$ eV, $E_{g\uparrow}=0.43$ eV, and $E_{g\downarrow}=1.24$ eV, respectively. The spin flip gaps for VB and CB are $\Delta^{VB}_{SF}=0.91$ eV and $\Delta^{CB}_{SF}=0.10$ eV, respectively. The band gap between the intrinsic electronic states for spin up and spin down are $E^{in}_{g\uparrow}=1.42$ eV and $E^{in}_{g\downarrow}=1.24$ eV, respectively.}
	\label{BS_DOS_1N_4N}
\end{figure*} 
The closest C$-$C distance of an intermolecular $\pi$ bonding between two adjacent C$_{60}$ molecules is 1.595 \AA, in good agreement with experimental findings (1.573 \AA).\cite{Meirzadeh2023} This very close intermolecular spacing ensures covalent bonding between C$_{60}$ molecules in graphullerene. 
\begin{figure*}[hbt]
	\begin{center}
		\includegraphics[width=7in]{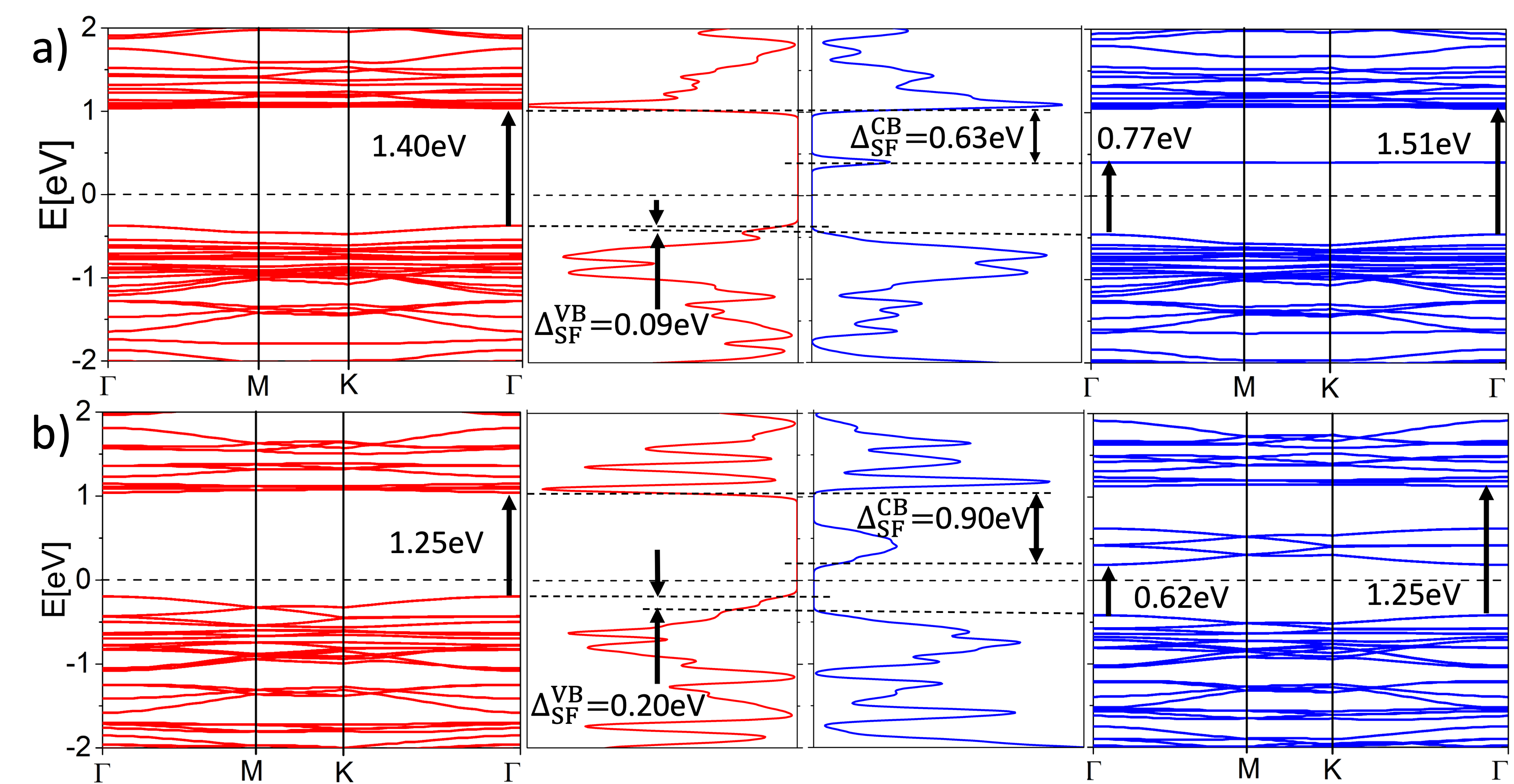}
	\end{center}
	\caption{a) The bandstructure and density of states for magnetic C$_{239}$B using spin-DFT. Red (blue) color denote spin up (down) electronic states. The charge, spin up, and spin down band gaps are $E_g=0.77$ eV, $E_{g\uparrow}=1.40$ eV, and $E_{g\downarrow}=0.77$ eV, respectively. The spin flip gaps for VB and CB are $\Delta^{VB}_{SF}=0.09$ eV and $\Delta^{CB}_{SF}=0.63$ eV, respectively. The band gaps between the intrinsic electronic states for spin up and spin down are $E^{in}_{g\uparrow}=1.40$ and $E^{in}_{g\downarrow}=1.51$ eV. b) The bandstructure and density of states for magnetic C$_{236}$B$_4$ using spin-DFT. Red (blue) color denotes spin up (down) electronic states. The charge, spin up, and spin down band gaps are $E_g=0.41$ eV, $E_{g\uparrow}=1.25$ eV, and $E_{g\downarrow}=0.62$ eV, respectively. The spin flip gaps for VB and CB are $\Delta^{VB}_{SF}=0.20$ eV and $\Delta^{CB}_{SF}=0.90$ eV, respectively. The band gaps between the intrinsic electronic states for spin up and spin down are $E^{in}_{g\uparrow}=1.25$ eV and $E^{in}_{g\downarrow}=1.25$ eV, respectively.}
	\label{BS_DOS_1B_4B}
\end{figure*} 
It is important to note that the covalent bonding between C$_{60}-$C$_{60}$ molecules in graphullerene is in stark contrast to the C$_{60}-$C$_{60}$ units in solid three-dimensional (3D) C$_{60}$, which crystallizes in fcc and sc structures, where the closest C$-$C distance between two C$_{60}$ molecules is 3.11 \AA. In solid C$_{60}$, fullerene molecules are considered as a smooth polarizable spherical shell, with two such shells attracting each other via van der Waals interaction. However, for solid C$_{60}$, it has been shown that coupling between different C$_{60}$ molecules through HP-edges and through the cornered atoms is energetically favored. A recent study on C$_{60}$ dimers also show that the edge bonding between C$-$C atoms on adjacent C$_{60}$ molecules and through pentagonal-hexagonal edge are energetically favored over several bonding schemes.\cite{Fullerene_Dimers} We see a similar trend in graphullerene. The bonding through pentagonal faces is not favored for closed packing such as hexagonal structures. Bonding through the HP-edge is energetically favored over the HH-edge due to electron-electron repulsion.   
 The separation between the C$_{60}-$C$_{60}$ molecule centers in graphullerene is obtained as 9.169 \AA, which is in good agreement with the previously reported values for C$_{60}$ dimers.\cite{Fullerene_Dimers} 

\paragraph*{Cohesive Energy:} 
The cohesive energy is a measure of the strength of the forces that bind atoms together in the solid state and is informative for studying phase stability. It is defined as the total energy of the constituent atoms minus the total energy of the compound. The cohesive energy of graphullerene is calculated to be 8.564 eV per carbon atom, indicating that graphullerene is stable at room temperature. The cohesive energy of graphullerene is comparable to the cohesive energy of solid fullerene (8.58 eV), in which fullerene crystallizes in the fcc crystal lattice, whereas graphullerene crystallizes in the 2D hcp crystal lattice.
 
\paragraph*{Symmetries and Bandstructure:}
The symmetry group of a molecule or solid material plays a crucial role in determining its molecular orbitals or electronic states and properties. Symmetry is a fundamental concept  and it provides a powerful framework for understanding the properties of molecules and materials. The isolated C$_{60}$ molecule has icosahedral symmetry, which is approximately maintained when the C$_{60}$ molecule is placed in the crystal environment of graphullerne. We observe only slight distortions. Numerical calculations show that the lowest unoccupied (LUMO) and the highest occupied (HOMO) molecular orbitals (MOs) of the C$_{60}$ molecule have primarily p$_z$ (also called $\pi$) character, where p$_z$ is a C 2p orbital pointing radially out of the nearly spherical molecule from the atom where it is centered. The HOMO and LUMO MOs of C$_{60}$ molecules transform according to the H$_u$ and T$_{1u}$ irreps of the I$_h$ symmetry group. The unit cell in graphullerene contains 2 pairs of C$_{60}$ molecules which are oriented differently, as shown in Fig.~\ref{structure_band_DOS}(a). Therefore, the unit cell of graphullerene is described by a much lower symmetry point group, i.e. C$_i$, containing only the identity and the inversion operator. C$_i$ has only two irreps, namely A$_g$ ($g$=gerade) and A$_u$ ($u$=ungerade) and are classified as even and odd with respect to the inversion operation. The presence of neighboring C$_{60}$ molecules in graphullerene weakly perturbs the graphullerence structure along the x and y directions, leading to the lifting of certain accidental degeneracies observed in the I$_h$ symmetry of the isolated C$_{60}$. Thus, the electronic states in graphullerene are even or odd with respect to the inversion operation, as shown in Fig.~1S in the Supplementary Information. 

\begin{figure*}[hbt]
	\begin{center}
		\includegraphics[width=7in]{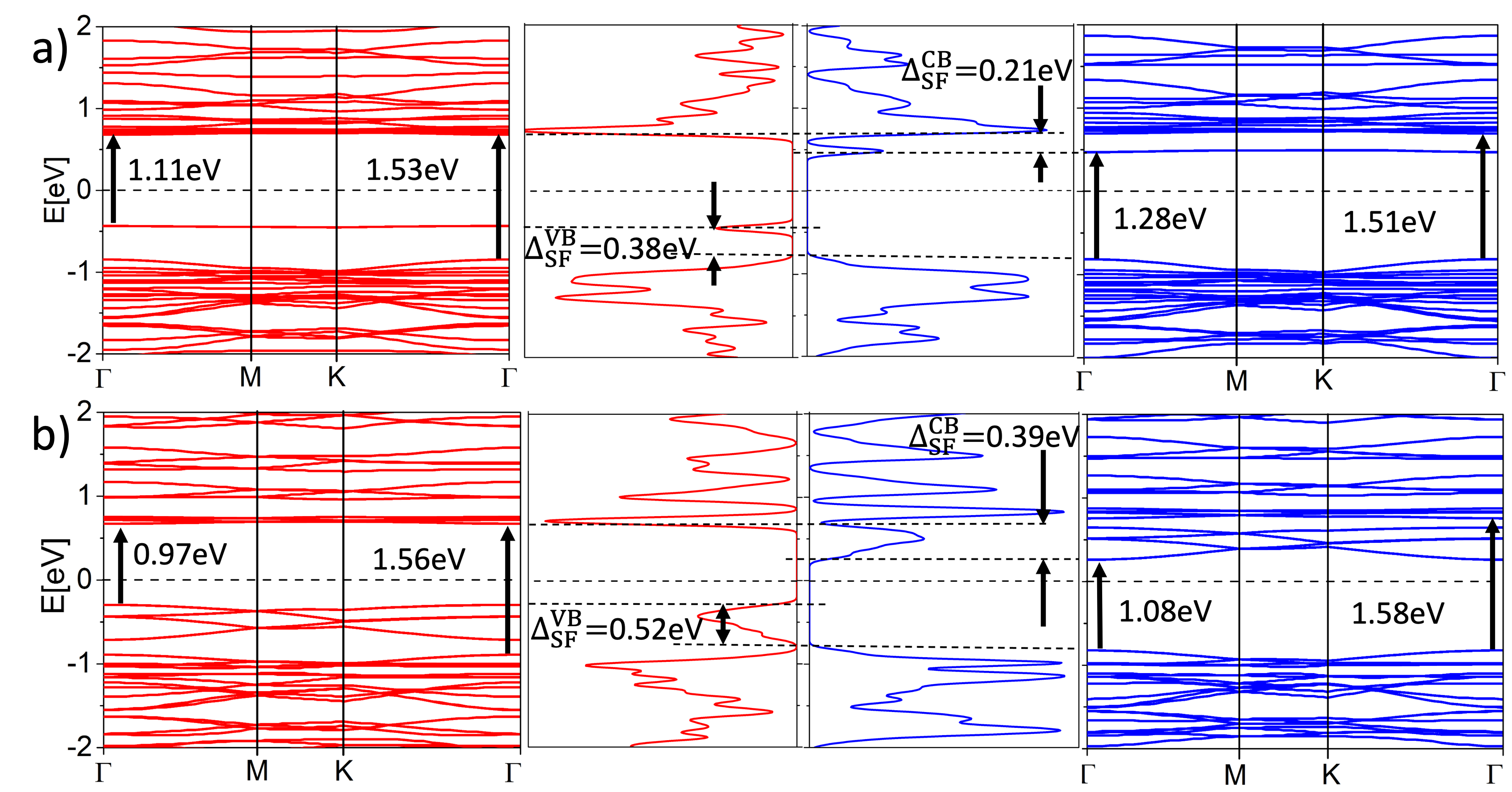}
	\end{center}
	\caption{a) The bandstructure and density of states for magnetic C$_{240}$H using spin-DFT. Red (blue) color is for spin up (down) electronic states. The charge, spin up, and spin down band gaps are $E_g=0.91$ eV, $E_{g\uparrow}=1.11$ eV, and $E_{g\downarrow}=1.28$ eV, respectively. The spin flip gaps for VB and CB are $\Delta^{VB}_{SF}=0.38$ eV and $\Delta^{CB}_{SF}=0.21$ eV, respectively. The band gaps between the intrinsic electronic states for spin up and spin down are $E^{in}_{g\uparrow}=1.53$ eV and $E^{in}_{g\downarrow}=1.51$ eV, respectively. b) The bandstructure and density of states for magnetic C$_{240}$H$_4$ using spin-DFT. Red (blue) color is for spin up (down) electronic states. Red (blue) color is for spin up (down) electronic states. The charge, spin up, and spin down band gaps are $E_g=0.56$ eV, $E_{g\uparrow}=0.97$ eV, and $E_{g\downarrow}=1.08$ eV, respectively. The spin flip gaps for VB and CB are $\Delta^{VB}_{SF}=0.52$ eV and $\Delta^{CB}_{SF}=0.39$ eV, respectively. The band gaps between the intrinsic electronic states for spin up and spin down are $E^{in}_{g\uparrow}=1.56$ eV and $E^{in}_{g\downarrow}=1.58$ eV, respectively.}
	\label{BS_DOS_1H_4H}
\end{figure*} 
The bandstructure is calculated along the $\Gamma -M-K-\Gamma$ path and shows that graphullerene has a direct bandgap of $E_g=1.48$ eV at the $\Gamma$-point and can therefore be classified as a semiconductor (Fig: \ref{structure_band_DOS}(c)). The bandgap of $E_g=1.48$ eV is in good agreement with the experimental value of 1.55 eV.\cite{Hou2022} Conduction band electronic states are nearly flat, which we attribute to the small hopping probability between the adjacent C$_{60}$ molecules and the resulting localization of the electronic states. In graphullerene two C$_{60}$ molecules are connected through sp$^3$ hybridized C$-$C bonds, which reduces the overlap of atomic orbitals and results in limited hopping of electrons between adjacent C$_{60}$ molecules. The cross linkage between two C$_{60}$ molecules in graphullerene typically consists of an sp$^3$-hybridized spacer. However, this spacer tends to interrupt the $\pi$-conjugation between the two cages, resulting in minimal or weak coupling between them. 

The MOs of a C$_{60}$ molecule can be mapped onto the electronic states of graphullerene, i.e. the irreps A$_g$, T$_{1g}$, T$_{2g}$, G$_g$, H$_g$ and A$_u$, T$_{1u}$, T$_{2u}$, G$_u$, H$_u$ of the I$_h$ group can be mapped on to A$_{g}$ and A$_{u}$ irreps of C$_i$ group. The compatibility relation is determined by the parity of the irreps. Bloch state analysis shows that valence and conduction bands of graphullerene are represented by A$_g$ and A$_u$ irreps of the C$_i$ group. The crystal field effects in graphullerene are substantial, which lead to the redistribution of electronic states and have a direct impact on the energy levels of MOs. Specifically, the crystal field effects cause the second highest occupied energy level HOMO-1 (Highest Occupied Molecular Orbital-1) of the isolated C$_{60}$ molecule to shift to a higher energy value, resulting in the formation of the valence band of graphullerene. 

\subsection{Impurity-engineered graphullerene}
Next, we consider the case of impurity-engineered graphullerenes (Fig.~\ref{structure_band_DOS})(b). As mentioned earlier, since each carbon atom in a C$_{60}$ molecule is identical and has four valence electrons, bonds to each of the three nearest-neighbor carbon atoms occur through sp$^2$ hybridization, while neighboring fullerenes are bonded through sp$^3$ hybridization in graphullerene. Since all the bonding requirements of the carbon atoms are satisfied, graphullerene is an intrinsic semiconductor with a band gap between the occupied valence band (VB) and unoccupied conduction band (CB), consistent with the observed electronic bandstructure (Fig.~\ref{structure_band_DOS}(c)). Impurities can greatly modify the properties of graphullerene in scientifically interesting and practically important ways. The electronic and optical properties of impurity-engineered isolated C$_{60}$ has been studies extensively. Nitrogen (N) and boron (B) substitutional impurities in isolated C$_{60}$ molecule has been successfully achieved experimentally, resulting in molecules with different impurity concentrations such as C$_{59}$N$^+$,\cite{C59N_doped, C59N_Doped_1} C$_{59}$HN,\cite{C59NH} C$_{60-m}$B$_m$ $(1\leq m\leq 6)$,\cite{C60_mB_m} C$_{60-n}$N$_n$ $(1\leq n\leq 4)$,\cite{C60_nN_n} and C$_{58}$BN.\cite{BN_C60} The impurities alter the charge distribution, electronic distribution, and energy gaps "$\Delta$" between the lowest and highest molecular orbitals of fullerenes.\cite{TOPC60} Similar results can be expected for graphullerene. The atomic size of N and B is relatively close to that of carbon, resulting in minimal strain or distortion. Also, the sp$^2$ hybridization of carbon atoms makes the atomic site highly reactive to adsorbent atoms such as H. Particular interest in H stems from the fact that H is always present in fullerene solids in noticeable amounts. H was detected in rather large concentration in ferromagnetic samples of pressure-polymerized fullerenes.\cite{Hyd_doping} Therefore, we consider six possible impurity engineering schemes, N impurities in graphullerene, i.e. C$_{239}$N and C$_{236}$N$_4$, B impurities in graphullerene, i.e. C$_{239}$B and C$_{236}$B$_4$, and H adsorbed on the surface of graphullerene, i.e. C$_{240}$H and C$_{240}$H$_4$. The formation energy for the different impurities and adsorbents are calculated by means of the relation
\begin{equation}\label{eq:FE}
E^f[\mathrm{C}_{240-n}\mathrm{X}_n]=E_{tot}[\mathrm{C}_{240-n}\mathrm{X}_n]-E_{tot}[\mathrm{C}_{240}]-\sum_{i}n_{i}\mu_{i}
\end{equation} 
$E_{tot}[\mathrm{C}_{240-n}\mathrm{X}_n]$ and $E_{tot}[\mathrm{C}_{240}]$ are the total energy of the system with and without the impurities, respectively, $n_i$ is the number of added ($n_i>0$) or removed ($n_i<0$) species of atoms during the formation of the impurity. 
$\mu_i$'s are chemical potentials of the C, N, B, and H atoms, which are estimated from their corresponding bulk forms. The calculated values of the formation energies are $E^f[\mathrm{C}_{239}\mathrm{N}=1.44$ eV, $E^f[\mathrm{C}_{236}\mathrm{N}_4=1.3$ eV, $E^f[\mathrm{C}_{239}\mathrm{B}=3.045049$ eV, $E^f[\mathrm{C}_{236}\mathrm{B}_4=2.97$ eV, $E^f[\mathrm{C}_{240}\mathrm{H}=-3.086362$ eV, and $E^f[\mathrm{C}_{240}\mathrm{H}_4=-3.25327$ eV. These correspond to an energy increase of 6 meV, 5.4 meV, 12.7 meV, 12.4 meV, -13 meV, and -13.3 meV per unit cell, respectively, indicating that the considered systems are thermodynamically stable. 

The results for bandstructure and density of states for different configurations i.e. C$_{239}$N, C$_{236}$N$_4$, C$_{239}$B, C$_{236}$B$_4$, C$_{240}$H, and C$_{240}$H$_4$ are shown in Figs.~\ref{BS_DOS_1N_4N}(a), \ref{BS_DOS_1N_4N}(b), \ref{BS_DOS_1B_4B}(a), \ref{BS_DOS_1B_4B}(b), \ref{BS_DOS_1H_4H}(a), and \ref{BS_DOS_1H_4H}(b), respectively. It becomes obvious that the N, B, and H impurities give rise to deep donor and deep acceptor levels.
To understand this, let us consider the thermal excitation of spin-up and spin-down charge carriers in impurity-engineered graphullerene. The electron-phonon interaction is spin-conserving if the material's spin-orbit coupling is weak.\cite{Han2014} This is the case for graphullerene, which makes graphullerene a promising candidate for spintronics applications. This also means that the spin band gaps play the major role because the spin is conserved during thermal excitation. In contrast to graphene, which is also useful for spintronics,\cite{Han2014} the spin-dependent band gaps in graphullerene are different for spin up and spin down. The relevant spin-dependent band gaps are shown in Table~\ref{bandgaps_spinflip}. The smallest spin-dependent band gap is 0.43 eV, corresponding to about 5000 K, which means that there are no thermally excited carriers in these half-semiconductors at room temperature. Thus, the donor and acceptor levels belong to deep donors and deep acceptors.

\section{Optical Properties of pristine and impurity-engineered graphullerene}

\subsection{Optical selection rules from symmetry}
The electric susceptibility for pristine graphullerene is shown ion Fig.~\ref{structure_band_DOS}(d) It is interesting to note that both the HOMO states, which transforms according to the irrep H$_u$ of the I$_h$ group, and the LUMO states, which transform according to the irrep T$_{1u}$ of the I$_h$ group, in isolated C$_{60}$ are odd under inversion. Therefore, the optical transitions are forbidden between the HOMO and LUMO states in isolated C$_{60}$.

In contrast to isolated C$_{60}$, pristine graphullerene has $C_{i}$ point group symmetry in the unit cell. The character table of $C_{i}$ is shown in Table~\ref{table_C_i} along with the irreps. The conduction band (CB) and valence band (VB) states near the $\Gamma$-point
transform according to $A_{g}$ and $A_{u}$ irreps of the $C_{i}$ group, respectively. Since the irreps $A_{g}$ and $A_{u}$ have opposite parity, optical transitions between the CB and VB states are allowed.

\begin{table}[h]%
\centering%
\begin{tabular}{ |c |c |c |c |c |c |c |c |c |}\hline
$C_{i}$    &    $E$    &    $i$ & linear functions, rotations & quadratic functions \\ 
\hline
$A_{g}$  & 1   & 1  & $R_x$, $R_y$, $R_z$  & $x^2$, $y^2$, $z^2$, $xy$, $xz$, $yz$ \\
$A_{u}$ & 1  & -1 & $x$, $y$, $z$ &  \\
\hline
\end{tabular}
\caption{Character table of the group $C_{i}$. $A_g$ and $A_u$ are the irreps of $C_{i}$.}
\label{table_C_i}
\end{table}

Let us analyze the optical selection rules for non-interacting electrons and holes in more detail.
The matrix elements of the susceptibility tensor ($i,j=x,y,z$) are determined by the Kubo–Greenwood formula, i.e.
\begin{equation}\label{eq:KGW}
\chi_{ij}(\omega)=\frac{e^2L}{4\pi\epsilon_0\hbar m^2_{e}V}\sum_{nm\bf{k}}\frac{(f_{m\bk}-f_{n\bk})\mu_{nm}^{i}(\bk)\mu_{mn}^{j}(\bk)}{\omega^2_{nm}({\bf{k}})[\omega_{nm}({\bf{k}})-\omega-i\Gamma/{\hbar}]}
\end{equation}  
where $\mu_{nm}^{j}=e\langle u_{n\bk}|r^{j}|u_{m\bk}\rangle$ is the electric dipole matrix, $V$ the volume of the crystal, $f_{nm\bk}=f_{n\bk}-f_{m\bk}$ is the difference between two Fermi functions, and $\Gamma=0.01$ eV the broadening. $\hbar\omega_{nm}(\bk)=E_{n\bk}-E_{m\bk}$ is the energy difference between the eigenenergies $E_{n\bk}$ and $E_{m\bk}$ of the Bloch states $\left|n\bk\right>$ and $\left|m\bk\right>$. $L=$25\AA\,is the length of the vacuum added along the z-axis.
\begin{figure*}[hbt]
	\begin{center}
		\includegraphics[width=7in]{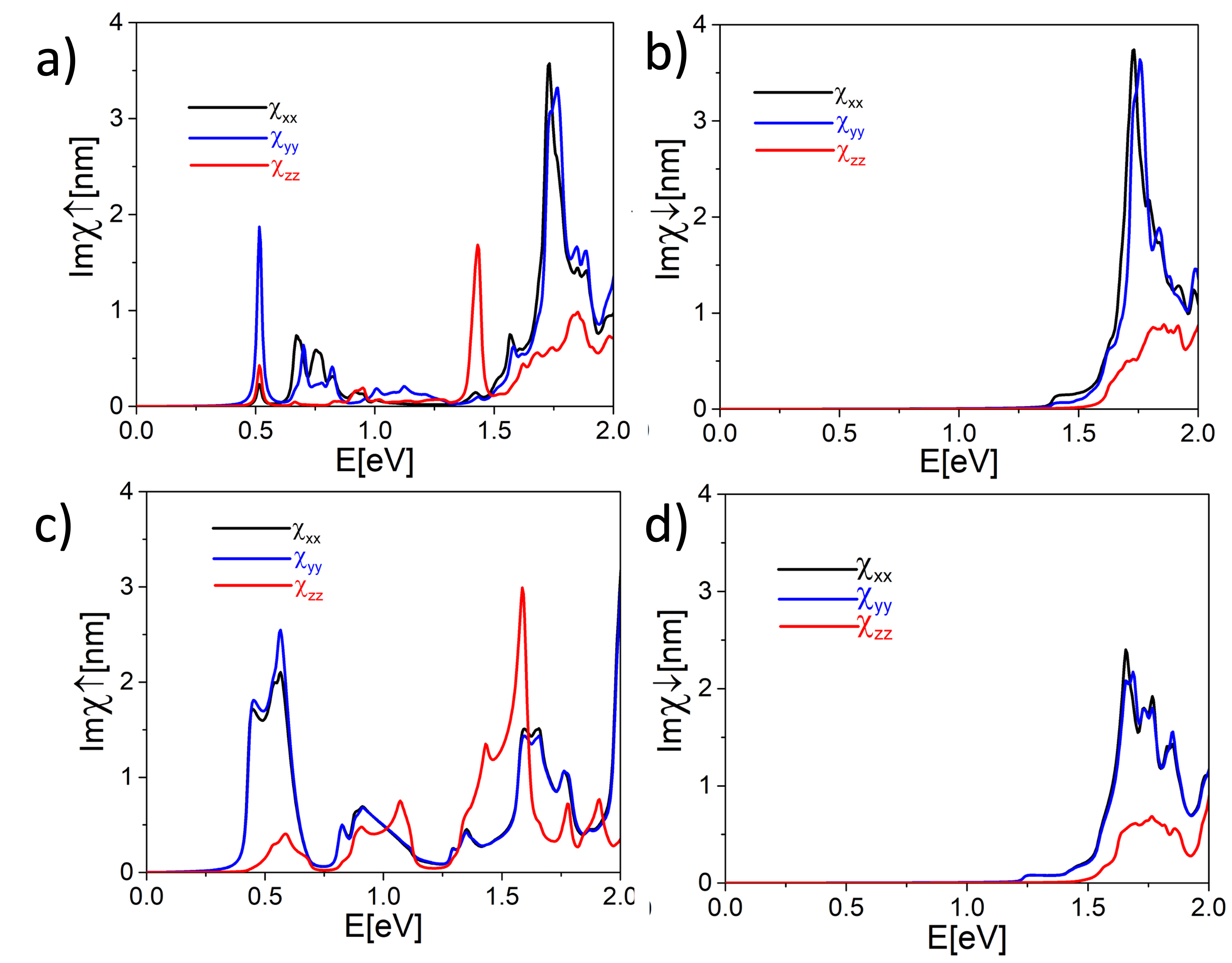}
	\end{center}
	\caption{Electric susceptibility for spin up $\chi\uparrow$ (a) and spin down $\chi\downarrow$ (b) for C$_{239}$N graphullerene. The absorption peaks in the subbandgap region shows the presence of localized impurity states. Electric susceptibility for spin up $\chi\uparrow$ (c) and spin down $\chi\downarrow$ (d) for C$_{236}$N$_4$ graphullerene. The absorption peaks in the subbandgap region shows the presence of localized impurity states.}
	\label{OS_1N_4N}
\end{figure*}
A general result from group theory states that an optical transition is allowed by symmetry only if the direct product of irreps $\Gamma(|u_{n\bk}\rangle)\otimes$ $\Gamma(\mu^j)$$\otimes\Gamma(|u_{m\bk}\rangle)$ contains $\Gamma(I)$ in its decomposition in terms of a direct sum. $\Gamma(I)$ denotes the irrep for the identity, i.e., $A_g$ for $C_{i}$. The resulting optical selection rules are shown in Table~\ref{table_C_i_selection_rules}. 

\begin{table}[h]
\begin{tabular}{|c |c |c |}\hline
$C_{i}$ & $A_{g}$ & $A_{u}$ \\
\hline
$A_{g}$ &  & $\sigma$, $\pi$ \\
\hline
 $A_{u}$ & $\sigma$, $\pi$  &  \\
\hline
\end{tabular}
\caption{Electric dipole selection rules for noninteracting electron-hole pairs in $C_{i}$ symmetry. $\sigma$ represents in-plane transitions while $\pi$ represents out-of-plane transitions.}
\label{table_C_i_selection_rules}
\end{table}

After identifying the CB with the $A_{g}$ irrep and the VB with the $A_{u}$ irrep from the Bloch states (see Supplementary Information), we find that the optical transition from VB to CB are allowed for both in-plane polarization, i.e. $x$- and $y$-polarization, corresponding to $\sigma$-transitions, and out-of-plane polarization, i.e. $z$-polarization, of the electric field, which corresponds to a $\pi$-transition. The optical selection rules for pristine graphullerene are summarized in Table~\ref{table_C_i_selection_rules}.

\begin{figure*}[hbt]
	\begin{center}
		\includegraphics[width=7in]{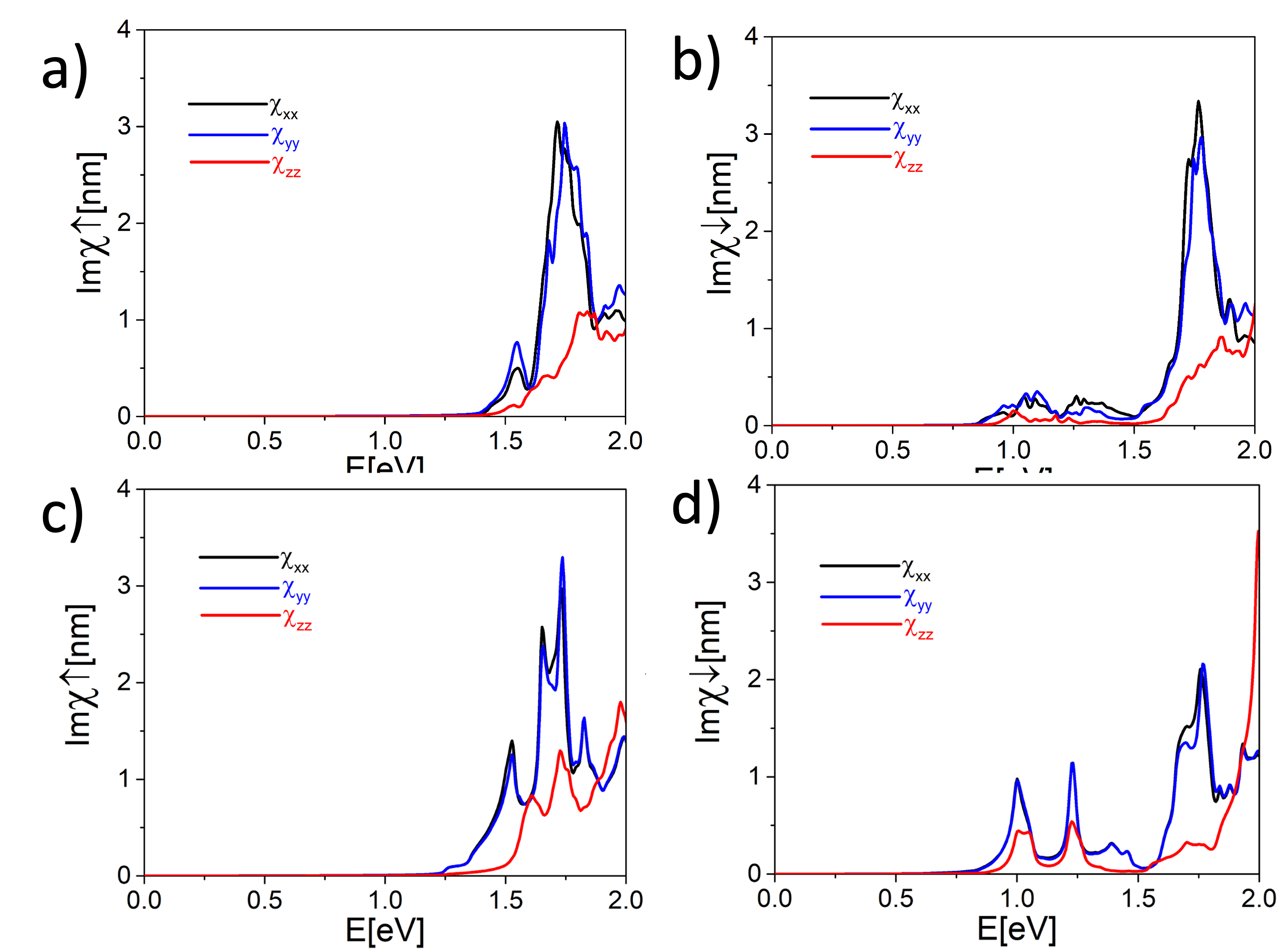}
	\end{center}
	\caption{Electric susceptibility for spin up $\chi\uparrow$ (a) and spin down $\chi\downarrow$ (b) for C$_{239}$B graphullerene. The absorption peaks in the sub bandgap region shows the presence of localized impurity states. Electric susceptibility for spin up $\chi\uparrow$ (c) and spin down $\chi\downarrow$ (d) for C$_{236}$B$_4$ graphullerene. The absorption peaks in the sub bandgap region shows the presence of localized impurity states.}
	\label{OS_1B_4B}
\end{figure*}

\begin{figure*}[hbt]
	\begin{center}
		\includegraphics[width=7in]{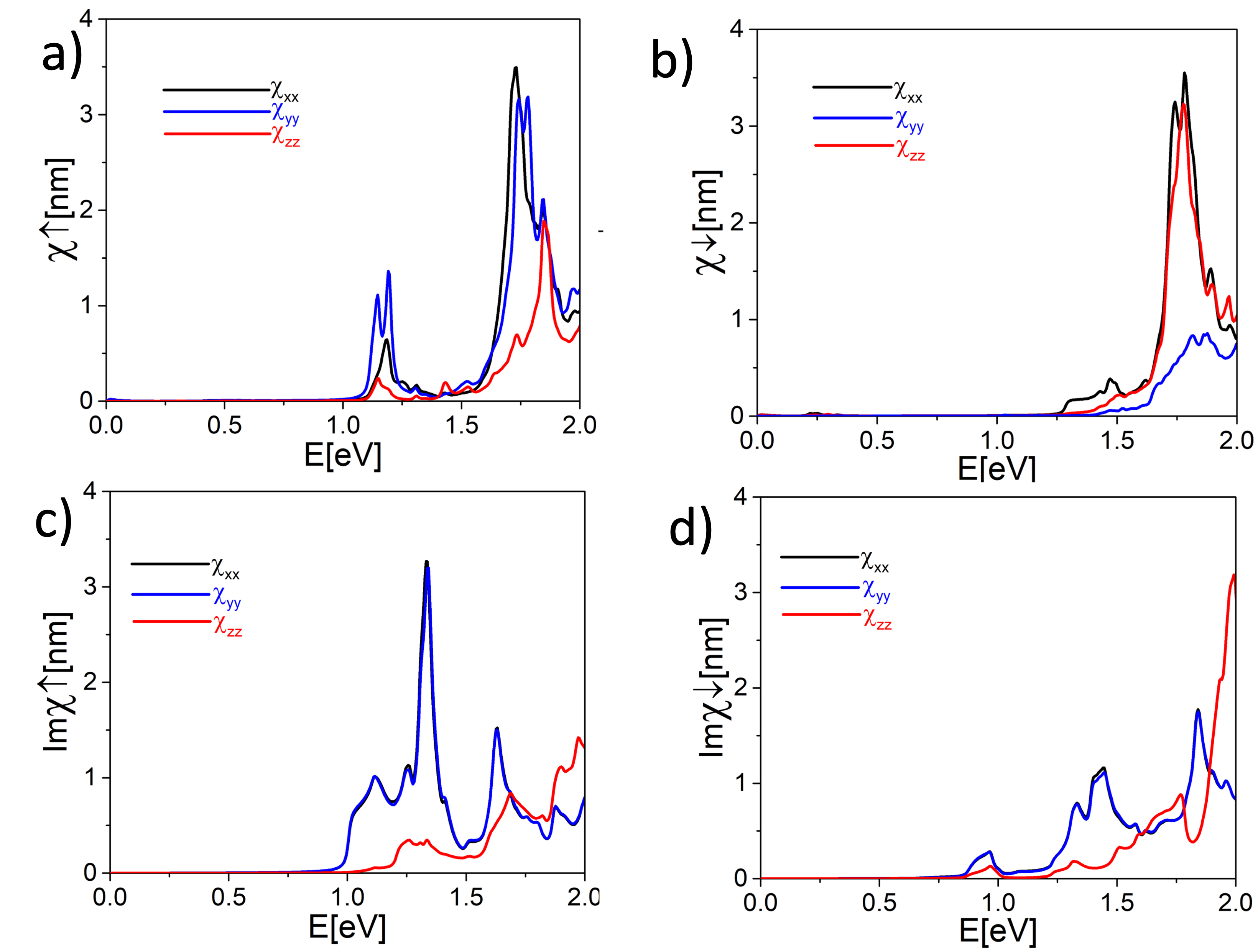}
	\end{center}
	\caption{Electric susceptibility for spin up $\chi\uparrow$ (a) and spin down $\chi\downarrow$ (b) for C$_{240}$H graphullerene. The absorption peaks in the sub-bandgap region shows the presence of localized impurity states. Electric susceptibility for spin up $\chi\uparrow$ (c) and spin down $\chi\downarrow$ (d) for C$_{240}$H$_4$ graphullerene. The absorption peaks in the sub-bandgap region shows the presence of localized impurity states.}
	\label{OS_1H_4H}
\end{figure*}

Interestingly, the presence of impurities breaks the C$_i$ symmetry. Therefore, impurity states belong to the C$_1$ symmetry, allowing optical transitions to and from states belonging to any irreps. This leads to absorption peaks in the sub-band gap region, as shown 
for C$_{239}$N, C$_{236}$N$_4$, C$_{239}$B, C$_{236}$B$_4$, C$_{240}$H, and C$_{240}$H$_4$ in Figs.~\ref{OS_1N_4N}(a), \ref{OS_1N_4N}(b), \ref{OS_1B_4B}(a), \ref{OS_1B_4B}(b), \ref{OS_1H_4H}(a), and \ref{OS_1H_4H}(b), respectively.

\subsection{Spin-dependent band gaps and spin-flip gaps}
In stark contrast to transition metal dichalcogenides (TMDs),\cite{Erementchouk2015,Khan2017,Khan2021,Khan2022} there are no strict optical selection rules distinguishing in- and out-of-plane polarization of the electric component of the electromagnetic field. However, there is a strict optical selection rule with respect to the spin of the band states, i.e. the initial and final states must have the same spin. In the case of pristine graphullerene the spin-up and spin-down bandstructures are identical, i.e. all the band states exhibit double spin degeneracy. However, in all versions of impurity-engineered graphullerene the spin-up and spin-down bandstructures differ substantially. This difference is typical of half-semiconductors, which can be characterized by three energy gaps, the band gap, the spin flip gap between valence band states, and the spin flip gap between conduction band states.\cite{Li2016} This lifting of the spin degeneracy is a direct consequence of time-reversal symmetry breaking, which in general lifts Kramers degeneracy. The origin is the magnetic dipole moments of the impurities (see Sec.~\ref{sec:magnetic_dipole}).

Specifically, for N-engineered graphullerene the spin-up and spin-down band gaps are both direct and at the $\Gamma$-point. The spin-up band gap is $E_{g\uparrow}=0.47$ eV, and the spin-down band gap is $E_{g\downarrow}=1.37$ eV. This means they differ by a factor of 2.9. The spin-flip gap between valence band states is $\Delta_{SF}^{VB}=0.84$ eV for 1N-engineered and $\Delta_{SF}^{VB}=0.91$ 4N-engineered graphullerene. The spin=flip gap between conduction band states is $\Delta_{SF}^{CB}=0.038$ eV for 1N-engineered and $\Delta_{SF}^{CB}=0.1$ eV 4N-engineered graphullerene.

For B-engineered graphullerene the spin-up and spin-down band gaps are both direct and at the $\Gamma$-point. However, the spin-up band gap is $E_{g\uparrow}=1.40$ eV, and the spin-down band gap is about $E_{g\downarrow}=0.86$ eV. This means they differ by a factor of 1.6. The spin-flip gap between valence band states is $\Delta_{SF}^{VB}=0.086$ eV for 1B-engineered and $\Delta_{SF}^{VB}=0.2$ 4B-engineered graphullerene. The spin=flip gap between conduction band states is $\Delta_{SF}^{CB}=0.063$ eV for 1B-engineered and $\Delta_{SF}^{CB}=0.90$ eV 4B-engineered graphullerene.

\begin{table}[h]
\begin{tabular}{|c |c |c |c |c |c|}\hline
Configuration & $E_g$ & $E_{g\uparrow}$ & $E_{g\downarrow}$ & $\Delta_{SF}^{VB}$ & $\Delta_{SF}^{CB}$  \\
\hline
C$_{240}$ & 1.48 & 1.48 & 1.48 & 0 & 0 \\
\hline
C$_{239}$N & 0.51  & 0.51 & 1.37 & 0.84 & 0.04 \\
\hline
C$_{236}$N$_4$ & 0.32 & 0.43 & 1.24 & 0.91 & 0.10 \\
\hline
C$_{239}$B & 0.77 & 1.40 & 0.77 & 0.09 & 0.63 \\
\hline
C$_{236}$B$_4$ & 0.41 & 1.25 & 0.62 & 0.20 & 0.90 \\
\hline
C$_{240}$H & 0.91 & 1.11 & 1.28 & 0.38 & 0.21 \\
\hline
C$_{240}$H$_4$ & 0.56 & 0.97 & 1.08 & 0.52 & 0.39 \\
\hline
\end{tabular}
\caption{Classification of half-semiconductors in terms of charge- and spin-dependent band gaps and spin-flip gaps of pristine and impurity-engineered graphullerene. For pristine graphullerene the spin flip gaps are zero, indicating that it is a standard semiconductor. Energy values are given in eV.}
\label{bandgaps_spinflip}
\end{table}


For H-adsorbed graphullerene the spin-up and spin-down band gaps are both direct and at the $\Gamma$-point. However, the spin-up band gap is $E_{g\uparrow}=1.12$ eV, and the spin-down band gap is $E_{g\downarrow}=1.29$ eV. This means they differ by a factor of 1.2. The spin-flip gap between valence band states is $\Delta_{SF}^{VB}=0.38$ eV for 1H-engineered and $\Delta_{SF}^{VB}=0.52$ 4H-engineered graphullerene. The spin=flip gap between conduction band states is $\Delta_{SF}^{CB}=0.21$ eV for 1H-engineered and $\Delta_{SF}^{CB}=0.39$ eV 4H-engineered graphullerene.

The various charge- and spin-dependent band gaps and spin-flip gaps of pristine and impurity-engineered graphullerene are summarized in Table~\ref{bandgaps_spinflip}.

\subsection{Exciton Binding Energy and Spin-Dependent Exciton States}
Nearly flat conduction bands in graphullerene correspond to localized states which result in very large exciton binding energies. The order of magnitude of the exciton binding energy can be estimated by using a modified Mott-Wannier model that accounts for the anisotropic dielectric properties and effective masses. The Wannier exciton binding energy is given by:
\begin{equation}
E_{WX} = \frac{\mu^* e^4}{2 (4 \pi \epsilon_0\epsilon_{r})^2 \hbar^2}
\end{equation}
where $\epsilon_{r}=\sqrt{\epsilon_{\parallel}\epsilon_{\perp}}$ is the relative permittivity of graphullerene and $\mu^{*}$ is the reduced mass, calculated as the geometric mean of the effective masses along three perpendicular axes:
\begin{equation}\label{eq:E_MASS}
\mu^{*} = \left( \frac{1}{\sqrt{\frac{1}{m_{ex} m_{hx}} + \frac{1}{m_{ey} m_{hy}} + \frac{1}{m_{ez} m_{hz}}}} \right)
\end{equation}
Here, $m_{e(h)i}$'s are the electron (hole) effective masses along the $i$th ($i=x, y, z$) axis in units of the rest electron mass. From DFT calculations, we find the reduced mass to be 0.6377 in units of bare electron mass, resulting in an exciton binding energy of $E_{WX}=3.37$ eV. This relatively large value of the exciton binding energy is attributed to the minimal screening effect due to the large reduced mass, reflected by the large density of states, and large effective mass. 

Another method to estimate the exciton binding energy is to use the formula for Frenkel excitons,\cite{pope1999electronic,knox1963excitons} i.e.
\begin{equation}
E_{FX} =\frac{{{e^2}}}{{4\pi {\epsilon_0}{\epsilon_r}a}},
\end{equation}
where $a$ is the effective Bohr radius of the exciton, which is the distance between the electron and the hole. We approximate $a$ by the diameter of a C$_{60}$ molecule, i.e. $a\approx 0.73$ nm. Then the Frenkel exciton binding energy becomes $E_{FX}=0.77$ eV. 

While both the Wannier and Frenkel exciton binding energies provide only rough estimates, the values inidcate that the exciton binding energy in graphullerene could be very large. More accurate calculations will be necessary in the future.

In addition, we emphasize the unique properties of spin-dependent exciton states in graphullerene. Our spin-dependent density functional theory (spin-DFT) calculations reveal that all versions of impurity-engineered graphullerene exhibit strongly spin-dependent band structures, resulting in distinct spin-polarized exciton states, which give rise to strict optical selection rules. Excitons with same spin in the conduction and valence band (opposite spins for electrons and  holes) result in bright excitons while excitons with opposite spins in the conduction and valence band (same spin for electrons and  holes) result in dark excitons.

These spin-dependent exciton states are of particular interest due to their potential applications in spintronic and opto-spintronic devices. The ability to generate and manipulate spin-polarized excitons using non-polarized light by tuning the energy of incident photons is a significant advancement. This capability can lead to novel optoelectronic devices that exploit the spin degrees of freedom, paving the way for advanced technologies in quantum computing and information storage.

The discovery of strongly spin-dependent exciton states in impurity-engineered graphullerene highlights its potential for groundbreaking applications in future electronic and photonic devices. The unique spin-dependent properties of these excitons offer a promising avenue for research and development in the rapidly evolving field of spintronics.

\section{Magnetic dipole moments of the impurities}
\label{sec:magnetic_dipole}
There is a great interest in the possible existence of magnetism in materials which have no transition metal or rare earth metal components. High temperature ferromagnetism has been reported in various carbon systems such as fullerenes \cite{Fullerene_mag_1, Fullerene_mag_2, makarova2006carbon} and graphene.\cite{graphene_mag_1} The magnetism involved in different allotropic forms of carbon is based on s and p electrons, rather than d or f as in transition metals or rare earths. By using spin-polarized DFT, we obtained the results for spin density, which is the difference in the up ($\rho_{\uparrow}$) and down ($\rho_{\downarrow}$) spin densities i.e. $\rho_{\uparrow}-\rho_{\downarrow}$ and magnetic moments $\mu$ are shown in Fig.~2S in the Supplementary Information. The sum of magnetic moments for singly impurity-engineered configurations such as C$_{239}$N, C$_{239}$B, C$_{240}$H is closed to unity, where as in case of tetra impurity-engineered configurations such as C$_{236}$N$_4$, C$_{236}$B$_4$, C$_{240}$H$_4$ the magnetic moment is closed to 4 in units of $\mu_B$.

\section{Conclusion}
By calculating the bandstructure of pristine and impurity-engineered ML graphullerene using spin-DFT, we have shown that graphullerene is a direct band gap semiconductor with a band gap $\sim$1.5 eV at the $\Gamma$ point, in very good agreement with experimental results. While pristine graphullerene is a regular semiconductor, we found that all versions of impurity-engineered graphullerene are half-semiconductors, with spin-dependent band gaps ranging from 0.43 eV ($\lambda\sim 2.9$ $\mu$m) to 1.5 eV ($\lambda\sim 820$ nm) for impurity concentrations of up to 1.7\%, thereby covering the near-infrared (NIR) and short-wavelength infrared (SWIR) regimes. 

In our detailed analysis, we discovered that impurity engineering with nitrogen, boron, and hydrogen in graphullerene lead to the formation of various half-semiconductors. This identification of half-semiconductors is particularly noteworthy because of the significant interest and potential applications it has garnered in recent times. The spin-dependent density functional theory (spin-DFT) calculations reveal that all versions of impurity-engineered graphullerene exhibit strongly spin-dependent band structures, making them candidates for spintronic and opto-spintronic devices. All the impurities give rise to a magnetic moment of approximately $\mu_B$ per impurity. The band gaps for the different spin species range from 0.43 eV to 1.5 eV, enabling the tuning of spin-polarized excitons and offering versatile applications in technology. Our results suggest that pristine and impurity-engineered graphullerene pave the way to develop carbon-based 2D semiconductor spintronic and opto-spintronic devices. The tunable band gap also holds great promise for solar energy harvesting, covering wavelengths from the infrared to the visible regime.

Our findings on the electronic and optical properties, especially the emergence of half-semiconductors in impurity-engineered graphullerene, position this material at the forefront of contemporary research in 2D materials. The ability to manipulate and tune the charge band gaps and the spin-dependent band gaps through selective impurity engineering underscores the versatility and potential of graphullerene for future technological advancements. 

Impurity-engineered graphullerene half-semiconductors with their deep donors and acceptors are versatile and find applications in a wide range of fields. Their unique properties and ability to introduce localized energy states within the band gap make them crucial for advancing various technologies. As we indicate, impurity-engineered graphullerene half-semiconductors are advanced materials that show promising applications for LEDs, photodetectors, high-temperature electronics, radiation detectors, thermoelectric materials, quantum computing devices, non-volatile memory devices, optoelectronic devices, sensors, MRAM, and nonlinear optical effects.

\section{Acknowledgments}
M. N. L. acknowledges support by the Air Force Office of Scientific Research (AFOSR) under award no. FA9550-23-1-0455 and support by the AFOSR under award no. FA9550-23-1-0472.


\section{References}

\bibliography{bibliography.bib}
\end{document}